\documentclass[preprint,3p,12pt]{elsarticle}

\usepackage{graphicx,subfigure}
\usepackage{amssymb}
\usepackage{amsmath,mathrsfs}
\usepackage{natbib}
\usepackage{color}
\usepackage{xcolor}
\usepackage{multirow}
\usepackage{soul}
 
\definecolor{codegreen}{rgb}{0,0.6,0}
\definecolor{codegray}{rgb}{0.5,0.5,0.5}
\definecolor{codepurple}{rgb}{0.58,0,0.82}
\definecolor{backcolour}{rgb}{0.95,0.95,0.92}

\newcommand{\hf}{\frac{1}{2}}
\newcommand{\PP}[2]{\frac{\partial#1}{\partial#2}}
\newcommand{\DD}[2]{\frac{\text{d}#1}{\text{d}#2}}
\newcommand{\ii}{\text{i}}
\newcommand{\re}[1]{#1_\text{R}}
\newcommand{\im}[1]{#1_\text{I}}
\newcommand{\tbd}[1]{#1_{\langle\cdot\rangle}}

\begin{document}

\begin{frontmatter}

\title{Integrating a ponderomotive guiding center algorithm into a quasi-static particle-in-cell code based on azimuthal mode decomposition}

\author[UCLAEE]{Fei Li} 
\ead{lifei11@ucla.edu}
\author[BNU]{Weiming An}
\author[UCLAPH]{Frank S. Tsung}
\author[UCLAPH]{Viktor K. Decyk}
\author[UCLAEE,UCLAPH]{Warren B. Mori}
\ead{mori@physics.ucla.edu}

\address[UCLAEE]{Department of Electrical Engineering, University of California Los Angeles, Los Angeles, CA 90095, USA}
\address[UCLAPH]{Department of Physics and Astronomy, University of California Los Angeles, Los Angeles, CA 90095, USA}
\address[BNU]{Department of Astronomy, Beijing Normal University, Beijing 100875, China}

\begin{abstract}
High fidelity modeling of plasma based acceleration (PBA) requires the use of three dimensional, fully nonlinear, and kinetic descriptions based on the particle-in-cell (PIC) method. In PBA an intense particle beam or laser (driver) propagates through a tenuous plasma whereby it excites a plasma wave wake.  Three-dimensional PIC algorithms based on the quasi-static approximation (QSA) have been successfully applied to efficiently model the interaction between relativistic charged particle beams and plasma. In a QSA PIC algorithm, the plasma response to a charged particle beam or laser driver is calculated based on forces from the driver and self-consistent forces from the QSA form of  Maxwell's equations. These fields are then used to advance the charged particle beam or laser forward by a large time step. Since the time step is not limited by the regular Courant-Friedrichs-Lewy (CFL) condition that constrains a standard 3D fully electromagnetic PIC code, a 3D QSA PIC code can achieve orders of magnitude speedup in performance. Recently, a new hybrid QSA PIC algorithm that combines another speedup technique known as an azimuthal Fourier decomposition  has been proposed and implemented. This hybrid algorithm decomposes the electromagnetic fields, charge and current density into azimuthal harmonics and only the Fourier coefficients need to be updated, which can reduce the algorithmic complexity of a 3D code to that of a 2D code. Modeling the laser-plasma interaction in a full 3D electromagnetic PIC algorithm is very computationally expensive due the enormous disparity of physical scales to be resolved. In the QSA the laser is modeled using the ponderomotive guiding center (PGC) approach.  
We describe how to implement a PGC algorithm compatible for the QSA PIC algorithms based on the azimuthal mode expansion. This algorithm permits time steps orders of magnitude larger than the cell size and it can be asynchronously parallelized. Details on how this is implemented into the QSA PIC code that utilizes an azimuthal mode expansion, QPAD, are also described. Benchmarks and comparisons between a fully 3D explicit PIC code (OSIRIS), as well as a few examples related to  laser wakefield acceleration, are presented.
\end{abstract}


\end{frontmatter}

\section{Introduction}

Laser wakefield acceleration (LWFAs) \cite{Joshi2003,Esarey2009} has attracted much interest and shown remarkable progress in the past four decades \cite{Joshi2020}. Due to the capability of producing large accelerating gradients and microscopic accelerating structures, LWFAs have immense promise for building compact and affordable accelerators for applications in high-energy physics and advanced radiation sources. In LWFA, when a relativistic laser pulse propagates through an under-dense plasma, the background electrons will be expelled sideways and forward  by the ponderomotive force of laser pulse after which they are attracted back towards the axis by the Coulomb force from the immobile ions. They then execute a multi-dimensional plasma oscillation creating a  wakefield. Charged particle beams riding at the correct phase of a wakefield (moving at close to the speed of light) will be continuously accelerated.

Modeling the complex physics in a LWFA requires self-consistently solving  Maxwell's equations and the relativistic equations of motion for plasma and trailing beam particles. Kinetic simulations based on the particle-in-cell (PIC) algorithm \cite{dawson1983,hockney1988,birdsall2005} have and continue to play a critical role in understanding and utilizing the LWFA processes. However, carrying out full 3D explicit start-to-end PIC simulations for LWFAs is very challenging mainly due the need to resolve the shortest spatial and temporal scales (typically the laser wavelength and period) and the huge disparity between these smallest physical scales and the acceleration length $L_\text{acc}$ ($\sim$mm to $\sim$m). We note that for some beam loading scenarios \cite{Wilks1987,Tzoufras2008} the trailing beam can be very narrow and this transverse scale is the smallest spatial scale needed to be resolved. For simplicity when discussing computational needs we assume that the smallest transverse scale is much larger than the laser wavelength.  Furthermore, there can be subtle numerical issues which can affect the fidelity of these simulations, e.g., numerical Cherenkov instability \cite{Godfrey2014,Xu2013,Lehe2013,Li2017}, numerical dispersion, and spurious forces \cite{Xu2020,Li2021a}. To mitigate or eliminate these deleterious effects often requires finer spatial-temporal resolution or more sophisticated numerical schemes, which inevitably increases the computational cost. For example, even with the today's most powerful supercomputers \cite{top500}, simulating a LWFA driven by a peta-watt laser pulse with ten-centimeter- to meter-class acceleration distance needs $\sim10^8$ CPU hours (assuming the use of the moving window \cite{Decker1994}) and therefore it still lacks feasibility with 3D explicit PIC codes. To achieve highly efficient modeling of LWFA processes, various speedup techniques based on reduced models have been developed, such as a Lorentz boosted frame technique \cite{Vay2007,Yu2016}, quasi-static approximation \cite{Mora1997}, ponderomotive guiding center model \cite{Mora1997, Gordon2000,Benedetti2017} and azimuthal Fourier decomposition \cite{Lifschitz2009,Davidson2015,Lehe2016}. In some cases these methods have been combined.

The ponderomotive guiding center (PGC) approximation is based on the idea of averaging the motion of a particle in the laser and wakefields over the laser frequency. This permits solving for the envelope of the laser and then using the laser envelope to calculate an average or ponderomotive force on the plasma electrons. This approximation works when there is no back-scattered light and there are no relativistic particles that reside within the laser pulse. With this time-averaging, the smallest physical scales needed to be resolved now become the spot size and wavelength of the wake which are typically the laser pulse spot size and duration. Therefore, the PGC approximation can provide speedups on the order of the factor $\sim(\omega_0/\omega_p)^2$ against the standard PIC simulations, where $\omega_0$ and $\omega_p$ are the laser and plasma frequency respectively. We typically have $\omega_0/\omega_p\gg1$ for LWFA problems of interests and thus the speedup provided by the PGC approximation is considerable. The PGC makes the computational needs of LWFA simulations similar to those of a PWFA per $c/\omega_p$ of driver propagation distance. The Lorentz boosted frame method can achieve speedups on the order of a factor of $\sim\gamma_f^2=(1-\beta_f^2)^{-2}$ by modeling the physics in a boosted frame moving at a speed of $\beta_f c$. In the boosted frame, the laser frequency is Doppler-shifted from $\omega_0$ to $\omega_0'=\omega_0/\gamma_f(1+\beta_f)$ and the plasma frequency $\omega_p \equiv \sqrt{\frac {4\pi e^2 n_p}{\gamma_f m_e}}$ is a Lorentz invariant. Thus, the cell size and time step used to resolve the laser wavelength are increased (the number of cells remains the same as the number of laser cycles is a Lorentz invariant). In this frame the plasma length (acceleration length) is also Lorentz contracted by a factor $\gamma_f^{-1}$.
The maximum speedup is achieved when $\omega_0'\sim\omega_p'$, which leads to a speedup that scales as $\gamma_f^2\sim(\omega_0/2\omega_p)^2$. The Lorentz boosted frame technique also requires that there is no back-scattered radiation and all forward moving modes have similar phase velocities. For both the PGC and Lorentz boosted frame methods it is assumed that the number of particles per cell is kept fixed when estimating the speedups. Therefore, the two approaches can provide similar speedups. 
Although the PGC can reduce the disparity between the minimum and maximum scales to be resolved, the use of a standard explicit PIC simulation is still subject to the limitation of time step size, which is a necessary condition to guarantee the numerical stability [known as the Courant–Friedrichs–Lewy (CFL) condition]. We note that even without the CFL constraint the time steps need to be small enough that particles do not move a distance approaching the smallest physical scale. The quasi-static approximation (QSA), initially presented as an analytical approach to study short-pulse laser interacting with plasma \cite{Mora1997}, was developed to treat the disparity in the physical scale between the evolution of lasers/charged particle beams and the plasma response. The QSA is essentially a multi-scale method that can separate the fast-varying plasma response, characterized by $\omega_p^{-1}$, from the slow-varying evolution of lasers/charged particle beams, characterized by Rayleigh length, $z_R\equiv \frac{\pi w_0^2}{\lambda_0}$ or betatron length, $\beta^*\equiv \frac{\sigma_0^2}{\epsilon}$, where $w_0$ and $\sigma_0$ are the spot sizes of the laser of beam respectively, $\lambda_0$ is the wavelength of the laser, and $\epsilon$ is the geometrical emittance of the beam. The QSA thus leads to speedup of $\beta^* \omega_p/c$ for PWFA simulations and an additional $z_R \omega_p/c$ speedup over the PGC method in standard PIC codes for LWFA simulations. We note that the PGC was first developed within the QSA \cite{Mora1997} and was then implemented into the fully electromagnetic PIC description \cite{Gordon2000}.

Numerically, a QSA PIC code \cite{Mora1997,Lotov2003,Huang2006,An2013,Mehrling2014,Li2021b,Wang2020,Diederichs2022} calculates the plasma response using QSA based field equations by assuming the laser or particle beam is static and then the resultant plasma-induced fields are used to advance the laser or particle beam by a large time step. The QSA field equations exclude radiative fields thus permitting time steps is excess of the CFL condition while still resolving the evolution of the driver.
For laser drivers, the QSA based PIC codes require a time averaging approach to model the laser and its associated ponderomotive force. This was implemented into a 2D r-z QSA based PIC code as described by \cite{Mora1997} and a 3D QSA based code as described by \cite{Huang2006,Wang2020}.

Another technique to boost the computational performance is known as the azimuthal Fourier decomposition \cite{Lifschitz2009,Davidson2015,Lehe2016} (also referred to as a quasi-3D or cylindrical mode expansion in the literature). In this technique, all the electromagnetic fields, and the charge and current density are expanded into a truncated Fourier series in $\phi$ (the azimuthal coordinate). The maximum number of Fourier harmonics that are kept is determined by the degree of asymmetry of the physical problem. This method turns the original 3D field updates into a series of 2D updates by only solving the complex Fourier coefficients defined on the 2D $r$-$z$ grid, and thus its computational complexity is greatly reduced. Thus it is a hybrid approach that uses a PIC description on an $r-z$ grid and a gridless description in $\phi$.  This approach has been successfully implemented into an explicit PIC framework \cite{Lifschitz2009,Davidson2015,Lehe2016}. 

Very recently, it was described how to merge the quasi-3D approach and the QSA. The result is the PIC code QPAD \cite{Li2021b} which is based on the numerical workflow of QuickPIC \cite{Huang2006,An2013}. It can enhance the computation efficiency by orders of magnitude compared to the full 3D standard and 3D QSA PIC codes. Implementing the PGC algorithm into a quasi-3D QSA based PIC code such as QPAD will enable simulating meter long LWFA simulations (that require peta-watt scale lasers) on small scale parallel computers (including desktop) while maintaining high fidelity so long as self-injection is not occurring and the laser does not overlap with the relativistic electron beam. The QSA and its neglect of radiative fields also has the advantage that it is more straightforward to handle small cell sizes in the transverse direction as can be needed for some nonlinear beam loading scenarios. 

In this paper, based on the existing PGC theoretical model \cite{Mora1997}, we extend the PGC algorithm from regular QSA \cite{Mora1997,Huang2006} and standard electromagnetic PIC codes \cite{Gordon2000,Benedetti2017} to a Fourier decomposed QSA code. In Ref. \cite{Benedetti2017} the PGC was implemented into an $r\text{-}\xi$ ($\xi=ct-z$) PIC code and the envelope solver included all derivatives. The implementation into QPAD will be described in detail. This algorithm consists of two components –- a Fourier decomposed finite difference solver and a QSA based particle pusher that includes the ponderomotive force. The proposed PGC laser solver is not constrained by the conventional CFL condition and is unconditionally stable for modeling lasers in vacuum. In the presence of plasma a large time step is still allowed. Moreover, the proposed algorithm is compatible with the pipelining technique in QPAD (based on ideas from QuickPIC \cite{Feng2009}) and thus can be executed asynchronously in parallel.

The paper is organized as follows: In Section \ref{sec:review}, we first briefly review the PGC theory, including the envelope equation for laser and the ponderomotive force for plasma particles. In Section \ref{sec:solver}, we describe the finite difference laser solver and provide a   proof of unconditional numerical stability. This is followed by the detailed description of the azimuthal Fourier decomposition form for the laser solver. Next,  the plasma susceptibility deposition scheme and the plasma particle pusher are presented in Section \ref{sec:chi_deposit} and \ref{sec:pusher}. In Section \ref{sec:example}, comparisons between QPAD \cite{Li2021b} and OSIRIS \cite{Fonseca2002,Hemker2015} for a  freely-drifting laser pulses in vacuum, standard LWFA driven by a regular Gaussian beam, and LWFA driven by a higher order Laguerre-Gaussian beam are presented. 

\section{Mathematical description of PGC model}
\label{sec:review}

This section will briefly review the main components of the PGC theoretical model \cite{Mora1997,Gordon2000}. The basic idea of PGC model is separating the physics into the rapidly varying components caused by the laser oscillation ($\tilde{\mathbf{a}}$) and the slowly varying components of the wake arising from averaging over the laser period ($\mathbf{A}$). Assuming the Coulomb gauge $\nabla\cdot(\mathbf{A}+\tilde{\mathbf{a}})=0$, the wave equation under the QSA for the vector potential is given by \cite{Mora1997}
\begin{equation}
\label{eq:wave_eq}
  \left(\nabla^2-\PP{^2}{t^2}\right)(\mathbf{A}+\tilde{\mathbf{a}}) = -\mathbf{J} - \tilde{\mathbf{j}} + \PP{}{t}\nabla\Phi
\end{equation}
where the uppercase symbols $\mathbf{A}$, $\Phi$ and $\mathbf{J}$ are the time-averaged vector potential, scalar potential and plasma current, while the lowercase symbols $\tilde{\mathbf{a}}$ and $\tilde{\mathbf{j}}$ are the vector potential and the current density associated with the fundamental frequency of the laser field. Note that the high frequency contribution to the scalar potential is dropped \cite{Mora1997}. Here and in the remainder of the paper,  normalized units are used. Length, time, mass and charge are normalized to $c\omega_p^{-1}$, $\omega_p^{-1}$, electron rest mass $m_e$, and electron charge $e$ respectively. The rapidly varying current density can be expressed in terms of the laser vector potential and relativistic factor [Eq. (A34) in Ref. \cite{Mora1997}] as
\begin{equation}
\label{eq:fast_jay}
  \tilde{\mathbf{j}} = -\tilde{\mathbf{a}}\sum_i\frac{q_i\rho_i}{\bar{\gamma_i} m_{0i}}
\end{equation}
where the $\rho_i$, $q_i$, $m_{0i}$ and $\bar{\gamma_i}$ are the charge density, charge, rest mass and averaged Lorentz factor of the {\it i}-th particle species.
The derivation of Eq. (\ref{eq:fast_jay}) uses the relationship between the rapidly varying current and plasma transverse momentum $\tilde{\mathbf{p}}_{\perp i}$, i.e., $\tilde{\mathbf{j}}=\sum_i \rho_i\tilde{\mathbf{p}}_{\perp i}/(\bar{\gamma}_i m_{0i})$, and the canonical momentum conservation $\tilde{\mathbf{p}}_{\perp i}=-q_i \tilde{\mathbf{a}}$. A more rigorous derivation can be found in the appendix of Ref. \cite{Mora1997}.
Separating the rapidly varying component of Eq. (\ref{eq:wave_eq}) out and combining with Eq. (\ref{eq:fast_jay}), the evolution of laser fields satisfies the following PDE
\begin{equation}
  \left(\nabla^2 - \PP{}{t^2} - \sum_i\frac{q_i\rho_i}{\bar{\gamma_i} m_{0i}}\right) \tilde{\mathbf{a}}=0.
\end{equation}
By assuming $\tilde{\mathbf{a}}$ can be expressed as the product of a rapidly varying phase term and a slowly varying complex envelope, i.e.,
\begin{equation}
  \tilde{\mathbf{a}} = \frac{a(\mathbf{x}_\perp, \xi, s)}{2}\exp(\ii k_0\xi)\hat{\mathbf{e}}_\perp + \text{c.c.}
\end{equation}
the evolution of the laser envelope $a$ satisfies
\begin{equation}
\label{eq:env_eq}
  2\PP{}{s}\left(\ii k_0+\PP{}{\xi}\right) a
  = (\nabla_\perp^2 +\chi) a
\end{equation}
where the Galilean transformation $\xi\equiv t-z$ ($z$ is the laser propagation direction), $s=t$ has been carried out, $k_0$ is the wavevector corresponding to $\omega_0$ and $\chi$, which describes the plasma response, is defined as
\begin{equation}
\label{eq:chi_def}
  \chi = -\sum_i\frac{q_i\rho_i}{\bar{\gamma}_i m_{0i}}.
\end{equation}
Mora and Antonsen \cite{Mora1997} showed that the time-averaged Lorentz factor is given by
\begin{equation}
  \bar{\gamma} = \left[1 + \bar{u}^2 + \left(\frac{q}{m_0}\right)^2\frac{|a|^2}{2} \right]^{1/2},
\end{equation}
where the $\bar{\mathbf{u}}$ is the time-averaged proper velocity.

The equation of motion for particles couples the evolution of the laser envelope and the plasma response together. The time-averaged equation of motion for plasma particles can be   expressed as
\begin{equation}
  \DD{\bar{\mathbf{u}}}{t} = \frac{q}{m_0}\left( \mathbf{E} + \frac{\bar{\mathbf{u}}}{\bar{\gamma}}\times\mathbf{B} - \frac{1}{4}\frac{q}{\bar{\gamma}m_0}\nabla|a|^2 \right)
\end{equation}
where $\mathbf{E}$ and $\mathbf{B}$ are the time-averaged electric and magnetic fields associated with the plasma response. The term having $\nabla|a|^2$ is the ponderomotive force from the laser. As long as the particle experiences enough laser oscillation cycles and the laser envelope changes little within a single cycle, the equation above is sufficiently accurate. Under the QSA, we have $\text{d}_t\rightarrow (1-\bar{v}_z)\text{d}_\xi$ and the time averaged equation of motion for a plasma particle becomes
\begin{equation}
\label{eq:motion_eq}
  \DD{\bar{\mathbf{u}}}{\xi} = \frac{q/m_0}{\bar{\gamma}-\bar{u}_z}\left( \bar{\gamma}\mathbf{E} + \bar{\mathbf{u}}\times\mathbf{B} - \frac{1}{4}\frac{q}{m_0}\nabla|a|^2 \right).
\end{equation}
The laser envelope equation [Eq. (\ref{eq:env_eq})], the equation of motion for plasma particles [Eq. (\ref{eq:motion_eq})] and the Maxwell's equations under QSA (see ref. \cite{An2013}) constitute a complete set of equations, and thus can be adopted as the working equations for a QSA simulation.

\section{Laser field solver}
\label{sec:solver}

\subsection{Finite difference solver}

Assuming that in a QSA simulation, the physical quantities associated with the plasma are defined on the integer time step $s=n\Delta_s$ while the laser field is defined on the half-integer time step $s=(n+\hf)\Delta_s$. The plasma and laser are then advanced forward in $s$ by the leapfrog method. Approximating $\partial_s a$ with a central difference operator, i.e., $\partial_s a\rightarrow (a^{n+\hf}-a^{n-\hf})/\Delta_s$, Eq. (\ref{eq:env_eq}) can be written as
\begin{equation}
  \label{eq:fd_direct}
  \left[\ii k_0 - \frac{1}{4}\Delta_s(\nabla_\perp^2+\chi^n)\right]a^{n+\hf} + \PP{a^{n+\hf}}{\xi} =
  \left[\ii k_0 + \frac{1}{4}\Delta_s(\nabla_\perp^2+\chi^n)\right]a^{n-\hf} + \PP{a^{n-\hf}}{\xi}.
\end{equation}

In a QSA PIC code, the asynchronous parallel algorithm (pipelining) \cite{Feng2009} is usually employed to improve the parallelism of the simulations. In this approach the quantities at smaller $\xi$ (the front of the driver) are known at larger values of $s$. Since using the regular central difference for $\partial_\xi a$ requires synchronous data communication among the MPI nodes distributed in the $\xi$-direction, then it is not possible to use a centered definition for $\partial_\xi a$. Instead, we choose a second-order backward difference operator to approximate $\partial_\xi a$, i.e.,
\begin{equation}
  \PP{a_j}{\xi} \rightarrow \text{D}_\xi a_j = \frac{1}{2\Delta_\xi}(3a_j-4a_{j-1}+a_{j-2}),
\end{equation}
where the index $j$ represents the field  defined at $\xi=j\Delta_\xi$. In this way, each MPI node in the $\xi$-direction only needs to receive data from the upstream node which keeps the value of $a_{j-1}$ and $a_{j-2}$ at $s+\Delta_s$ and $s$ at the MPI boundary (at index j). 

Another complication arises from the $\chi^n a^{n+\hf}$ term on the LHS of Eq. (\ref{eq:fd_direct}) for the quasi-3D algorithm. The value of $\chi^n$ depends on position, thus the differential equation for $a$ is linear but with "spatially" dependent (non constant) coefficients. In the quasi-3D algorithm, the fields are expanded into azimuthal harmonics, i.e., Fourier modes. The presence of the spatially varying $\chi^n a^{n+\hf}$ term thus leads to cross-product terms between different azimuthal modes of $\chi^n$ and $a^{n+\hf}$. This makes the equations for the different azimuthal modes of $a^{n+\hf}$ coupled together, adding extra difficulties to the implementation compared with a fully finite difference method. 
However, as we describe shortly if the couplings are carefully accounted for then the $s$ advance for each mode requires the inversion of a single constant coefficient matrix.

Therefore, to address both issues we have constructed the following iterative formula for Eq. (\ref{eq:fd_direct})
\begin{equation}
\begin{aligned}
  \label{eq:fd_iter}
  &\left(\ii k_0 - \frac{1}{4}\Delta_s\nabla_\perp^2 + \frac{3}{2\Delta_\xi}\right)a_j^{n+\hf,l} \\
  &= \frac{1}{4}\Delta_s\chi_j^n a_j^{n+\hf,l-1} + \frac{1}{2\Delta_\xi}(4a_{j-1}^{n+\hf} - a_{j-2}^{n+\hf}) +
  \left[\ii k_0 + \frac{1}{4}\Delta_s(\nabla_\perp^2+\chi_j^n)\right]a_j^{n-\hf} + \text{D}_\xi a_j^{n-\hf} 
\end{aligned}
\end{equation}
Here $l$ is the iteration step and $a^{n+\hf,l}\rightarrow a^{n+\hf}$ when $l\rightarrow\infty$ if the iteration is converged. For this iterative finite difference equation, the LHS (which includes the unknown values of the laser field for the next iteration) only contains finite difference operators that have constant coefficients.
We point out that Eq. \ref{eq:fd_iter} is applicable to the full 3D Cartesian, full 3D cylindrical, and the quasi-3D descriptions. The primary differences between them lies in the specific form of the transverse Laplacian $\nabla_\perp^2$ and the calculation of the $\chi^n a^{n+\hf}$ term. For the full 3D Cartesian coordinate system, $\nabla_\perp^2$ can be approximated using the classical 5-point formula and thus the linear system solvers like fast FFT and multigrid method are the usual choice. For the quasi-3D algorithm, as we will describe in section \ref{sec:afd} the $\nabla_\perp^2$ is split into multiple 1D operators in $r$, and we need to calculate the cross-product terms of $\chi^n a^{n+\hf}$ as stated before.

\subsection{Convergence and stability}

It is important to understand the stability and the rate of convergence of the proposed iteration loop. In the case where the plasma is absent, i.e., $\chi=0$, there is no need to iterate over Eq. (\ref{eq:fd_iter}) and the finite difference wave equation reduces to the regular Crank-Nicolson (CN) method which is unconditionally stable \cite{Thomas2013}. 

When plasma is present ($\chi\neq0$), the convergence condition of the iteration over $l$ can be determined when $\chi^n$ is not spatially dependent. It can be proven (see \ref{sec:app_iter_conv_cond}) that the convergence condition in this situation is
\begin{equation}
\label{eq:conv_cond}
  k_0^2\Delta_\xi^2 + \frac{9}{4} > \frac{1}{16}|\chi^n|^2\Delta_s^2\Delta_\xi^2.
\end{equation}
For typical simulation scenarios, we have $\chi\sim O(1)$, $\Delta_\xi\sim O(0.1)$ and the validity of the PGC model requires $k_0\gg1$. This allows for a large $\Delta_s$ [e.g. $\sim O(1)$ to $\sim O(10)$] according to the convergence condition [Eq. (\ref{eq:conv_cond})].
In reality $\chi$ is not spatially uniform making finding an analytical convergence condition difficult. We can however still qualitatively determine $\Delta_s$ through Eq. (\ref{eq:conv_cond}) by replacing $\chi^n$ with the maximum $\chi$ on the grid.

For an infinite number of iterations ($l\rightarrow\infty$) and for a spatially independent $\chi$, it can be shown that the proposed scheme is stable without any restrictions for $\Delta_s$. However, for the more relevant scenario where $l$ is finite and $\chi$ is not spatially uniform, mathematically analyzing the numerical stability becomes very difficult. Nevertheless, numerous tests indicates that the proposed scheme is numerically stable even for a large $\Delta_s$ [$\sim O(10)$] for a wide variety of LWFA-relevant problems.

\subsection{Azimuthal Fourier decomposition}
\label{sec:afd}

In order to make the proposed finite difference scheme applicable to QPAD, we need to further expand Eq. (\ref{eq:fd_iter}) it into azimuthal modes and derive the finite difference equations for each azimuthal harmonic. The complex-valued iterative equation [Eq. (\ref{eq:fd_iter})] is thus split into two coupled real-valued equations corresponding to the real and imaginary parts respectively. The real and imaginary parts of $a$, i.e., $\re{a}$ and $\im{a}$, are expanded using the following azimuthal Fourier series,
\begin{equation}
  \tbd{a} = \tbd{a}^0 + 2\sum_{m=1}^{+\infty}\mathfrak{Re}\{\tbd{a}^m\}\cos(m\phi) - 2\sum_{m=1}^{+\infty}\mathfrak{Im}\{\tbd{a}^m\}\sin(m\phi)
\end{equation}
where $\langle\cdot\rangle$ denotes ``R'' or ``I'' and $m$ is the harmonics number. The complex Fourier coefficients $\re{a}^m$ and $\im{a}^m$ obey the following equations
\begin{align}
  \label{eq:eqn_re_rz}
  \left(-\frac{1}{4}\Delta_s\Delta_m + \frac{3}{2\Delta_\xi}\right) a_{\text{R},j}^{m,l} - k_0 a_{\text{I},j}^{m,l} &= \frac{1}{4}\Delta_s (\chi_j a_{\text{R},j})^{m,l-1} + \frac{1}{2\Delta_\xi} (4a_{\text{R},j-1}^m - a_{\text{R},j-2}^m ) + S_{\text{R},j}^m, \\
  \label{eq:eqn_im_rz}
  \left(-\frac{1}{4}\Delta_s\Delta_m + \frac{3}{2\Delta_\xi}\right) a_{\text{I},j}^{m,l} + k_0 a_{\text{R},j}^{m,l} &= \frac{1}{4}\Delta_s (\chi_j a_{\text{I},j})^{m,l-1} + \frac{1}{2\Delta_\xi} (4a_{\text{I},j-1}^m - a_{\text{I},j-2}^m ) + S_{\text{I},j}^m,
\end{align}
where the Laplacian for $m$-th harmonic $\Delta_m\equiv \frac{1}{r}\PP{}{r}\left(r\PP{}{r}\right)-\frac{m^2}{r^2}$. Without introducing ambiguity, we have left out the time indices to simplify the denotations. Here, the symbols $\re{a}$, $\im{a}$ and $\chi$ explicitly appearing in Eq. (\ref{eq:eqn_re_rz}) and (\ref{eq:eqn_im_rz}) are defined at $s=(n+\hf)\Delta_s$ while those included in $\re{S}^m$ and $\im{S}^m$, given by
\begin{align}
  \label{eq:SR}
  S_{\text{R},j}^m &= \frac{1}{4}\Delta_s (\Delta_m a_{\text{R},j}^m + (\chi_j a_{\text{R},j})^m) - k_0 a_{\text{I},j}^m + \text{D}_\xi a_{\text{R},j}^m \\
  \label{eq:SI}
  S_{\text{I},j}^m &= \frac{1}{4}\Delta_s (\Delta_m a_{\text{I},j}^m + (\chi_j a_{\text{I},j})^m) + k_0 a_{\text{R},j}^m + \text{D}_\xi a_{\text{I},j}^m,
\end{align}
are defined at $s=(n-\hf)\Delta_s$. The nonlinear terms in Eqs. (\ref{eq:eqn_re_rz}) to (\ref{eq:SI}) are calculated via the following truncated series
\begin{equation}
  (\chi a_{\langle\cdot\rangle})^m = \sum_{k=m-M}^M \chi^k a_{\langle\cdot\rangle}^{m-k}
\end{equation}
where $M$ is the maximum harmonic number involved in the simulation.

Using a 3-point discretization, the $\Delta_m$ operator with second-order precision can be written as
\begin{equation}
  \Delta_m \tbd{a}^m \rightarrow \beta_i^- a_{\langle\cdot\rangle,i-1}^m - \alpha_i a_{\langle\cdot\rangle,i}^m + \beta_i^+ a_{\langle\cdot\rangle,i+1}^m
\end{equation}
where
\begin{equation}
  \beta_i^\pm = \frac{1}{\Delta_r^2}\left(1 \pm \frac{1}{2i}\right), \quad
  \alpha_i = \frac{1}{\Delta_r^2}\left(2 + \frac{m^2}{i^2}\right).
\end{equation}
The subscript $i$ of $\tbd{a}^m$ means it is defined at the radial position $r=i\Delta_r$. The coupled equations (\ref{eq:eqn_re_rz}) and (\ref{eq:eqn_im_rz}) can be recast into a penta-diagonal linear system $PX=\text{RHS}$, where the unknown vector $X$ is constructed by alternatively placing $\re{a}$ and $\im{a}$
\begin{equation}
  X=(\cdots,a_{\text{R},i}^m,~a_{\text{I},i}^m,~a_{\text{R},i+1}^m,~a_{\text{I},i+1}^m,\cdots).
\end{equation}
The penta-diagonal coefficient matrix $P$ is given by
\begin{equation}
  \begin{pmatrix}
    \ddots & \ddots & \ddots & \ddots & \ddots & & & & & \\
    & B_i^- & 0 & A_i & -k_0 & B_i^+ & & & &\\
    & & B_i^- & k_0 & A_i & 0 & B_i^+ & & &\\
    & & & B_{i+1}^- & 0 & A_{i+1} & -k_0 & B_{i+1}^+ & & \\
    & & & & B_{i+1}^- & k_0 & A_{i+1} & 0 & B_{i+1}^+ & \\
    & & & & & \ddots & \ddots & \ddots & \ddots & \ddots \\
  \end{pmatrix}
\end{equation}
where $B_i^\pm=\frac{1}{4}\Delta_s\beta_i^\pm$ and $A_i=-\frac{1}{4}\Delta_s\alpha_i + \frac{3}{2\Delta_\xi}$. This linear system can be efficiently solved  using the cyclic reduction method \cite{Levit1989}.

\subsection{Boundary conditions}

The boundary conditions at $r=0$ is different for the $m=0$ and $m>0$ modes. For $m=0$ mode, $\partial_r \tbd{a}^0|_{r=0}=0$ and the Laplacian $\Delta_0=\frac{1}{r}\PP{}{r}\left(r\PP{}{r}\right) \rightarrow 2\PP{^2}{r^2}$ when $r\rightarrow 0$. The former indicates that $a_{\langle\cdot\rangle,i=-1}=a_{\langle\cdot\rangle,i=1}$. In this situation, the stencil coefficients for $i=0$ becomes $\beta_0^- = 0$, $\beta_0^+=\alpha_0=2/\Delta_r^2$. For $m>0$ modes, the symmetry requires $\tbd{a}^m=0$ at $r=0$, and thus $a_{\langle\cdot\rangle,i>0}^m$ can be solved without changing the stencil coefficients.

A Dirichlet boundary condition is used for the laser envelope of all the harmonics at $r=r_\text{max}$. As long as the laser envelope does not extend to the boundaries, a Dirichlet boundary condition can work properly without introducing unphysical results.

\section{Plasma susceptibility deposition}
\label{sec:chi_deposit}

In order to solve the laser envelope equations  for each harmonic [Eqs. (\ref{eq:eqn_re_rz}) to (\ref{eq:SI})], the plasma susceptibility $\chi^m$ needs to be deposited onto the grid based  on   the particle information, i.e., charge, position, momentum and mass. The deposition scheme for $\chi^m$ basically follows that used for the source terms (charge density, current density, etc.) as described in Section 3.4 of the original QPAD paper \cite{Li2021b}. A more convenient form for $\chi$ [Eq. (\ref{eq:chi_def})] when dealing with particle data is
\begin{equation}
  \chi = -\frac{q}{m_0}\frac{\rho-J_z}{\bar{\gamma}-\bar{u}_z},
\end{equation}
which is obtained by multiplying both the numerator and denominator of Eq. (\ref{eq:chi_def}) by $(1-\bar{v}_z)$.

We start from the general formula for depositing $\chi$ as a sum over particles,
\begin{equation}
  \chi = \frac{1}{\text{vol.}} \frac{q}{m_0}\sum_i \frac{q_i}{\bar{\gamma}_i-\bar{u}_{z,i}} \frac{1}{r}S_r(r-r_i)S_\phi(\phi-\phi_i),
\end{equation}
where $q_i$, $r_i$, $\phi_i$, $\bar{\gamma}_i$ and $\bar{u}_{z,i}$ are the charge, radial position, azimuthal angle, time-averaged Lorentz factor and longitudinal proper velocity of the $i$-th particle. The particle shape function $\frac{1}{r}S_r(r-r_i)S_\phi(\phi-\phi_i)$ is used to interpolate the particle information at position $(r_i,~\phi_i)$ onto the grid position $(r,~\phi)$. Both $S_r$ and $S_\phi$ should satisfy the normalization conditions $\int dr S_r=1$ and $\int d\phi S_\phi=1$. Since QPAD uses the gridless description in $\phi$, we need to expand the azimuthal shape function $S_\phi$ into azimuthal harmonics,
\begin{equation}
  S_\phi(\phi-\phi_i) = \sum_m S_\phi^m(\phi_i) e^{\ii m\phi}
\end{equation}
where the Fourier coefficient is expressed as
\begin{equation}
  S_\phi^m(\phi_i) = \frac{1}{2\pi}\int_0^{2\pi}d\phi' S_\phi(\phi'-\phi_i)e^{-\ii m\phi'}.
\end{equation}
In QPAD we adopt the Dirac delta function as the particle shape function in $\phi$, i.e., $S_\phi(\phi-\phi_i)=\delta(\phi-\phi_i)$ and thus $S_\phi^m(\phi_i)=e^{-\ii m\phi_i}/(2\pi)$. Therefore, the deposition formula for the $m$-th harmonic of $\chi$ can be written as
\begin{equation}
  \chi^m = \frac{1}{\text{vol.}} \frac{q}{m_0}\sum_i \frac{q_i}{\bar{\gamma}_i-\bar{u}_{z,i}} \frac{1}{2\pi r}S_r(r-r_i)e^{-\ii m\phi_i}.
\end{equation}
In practice, it is not necessary to calculate each harmonic but only the $m=0$ mode from each particle. The contribution to any $m>0$ mode from an individual particle can be calculated from the $m=0$ contribution by simply multiplying a constant phase factor, i.e., through the relation $\chi^m=\chi^0 e^{-\ii m\phi_i}$ or recursively through $\chi^m=\chi^{m-1} e^{-\ii \phi_i}$.

\section{Particle pusher}
\label{sec:pusher}

In QPAD, all the physical quantities associated with the plasma are defined on the integer time step $s=n\Delta_s$. Within a time step, the shape of the laser or particle beam is assumed unchanged, and the ponderomotive force or the Coulomb like force of a drive or trailing beam is used to update the plasma response. The quantities associated with plasma, including the information of plasma particles, the charge and current density and the plasma-induced electromagnetic fields, are updated from the front to the end of the moving window for each transverse slice. The particle information and the fields are solved using Eq. (\ref{eq:motion_eq}) and the Maxwell's equations with QSA described in ref. \cite{An2013}. Before numerically integrating Eq. (\ref{eq:motion_eq}), one needs to interpolate $\mathbf{E}$, $\mathbf{B}$ and $\nabla|a|^2$ from the grid points onto the position of each particle. Instead of calculating $\nabla|a|^2$ on the grid points first and then interpolating onto the particles' positions, we choose to interpolate $a$ and $\nabla a$ separately. The $|a|^2$ term in $\bar{\gamma}$ for an individual particle is calculated from the interpolated $a$, and the $\nabla |a|^2$ at the location for an individual particles is calculated via $\nabla|a|^2=2\mathfrak{Re}\{a^*\nabla a\}$. This approach will reduce the number of floating point operations especially when there are many Fourier harmonics. The formula for interpolating $a$ from the grid points onto the $i$-th particle position $(r_i,~\phi_i)$ can be derived as follows
\begin{equation}
\begin{split}
  a(r_i,\phi_i) &= \int \text{d}r \int \text{d}\phi\ a(r,\phi)S_r(r-r_i)S_\phi(\phi-\phi_i) \\
  &= \frac{1}{2\pi}\int\text{d}r \int\text{d}\phi\ \sum_m a^m(r)e^{\ii m\phi}\sum_{m'}e^{\ii m'\phi_i}e^{-\ii m'\phi} S_r(r-r_i) \\
  &= \sum_m\int\text{d}r\ a^m(r)e^{\ii m\phi_i} S_r(r-r_i).
\end{split}
\end{equation}
The interpolation of $\mathbf{E}$ and $\mathbf{B}$ can be done in a similar way. Since $\nabla_\perp a^m=\PP{a^m}{r}\mathbf{e}_r + \frac{\ii m}{r}a^m \mathbf{e}_\phi$, the formula for interpolating $\nabla a$ is given by
\begin{equation}
  \nabla a(r_i,\phi_i) = \sum_m \int\text{d}r\ \left(\PP{a^m}{r}\mathbf{e}_r + 
  \frac{\ii m}{r}a^m\mathbf{e}_\phi + \PP{a^m}{\xi}\mathbf{e}_z\right)e^{\ii m\phi_i}S_r(r-r_i).
\end{equation}

In QPAD, the plasma particle positions are defined on the integer grid points in $\xi$, i.e., $\xi=j\Delta_\xi$, so that the fields and ponderomotive force felt by particles are also located at $\xi=j\Delta_\xi$. The particle momenta are defined on the half-integer grid points, i.e., $\xi=(j+\hf)\Delta_\xi$. After calculating the fields and ponderomotive force at the particle positions, we can numerically integrate Eq. (\ref{eq:motion_eq}) using the Boris method \cite{Boris1970}. It should be noted that when advancing $\bar{\mathbf{u}}^{j-\hf}$ to $\bar{\mathbf{u}}^{j+\hf}$, one needs to know the values of $\bar{\gamma}$ and $\bar{\gamma}-\bar{u}_z$ at $\xi=j\Delta_\xi$ to make the Boris method applicable. These two values can be obtained during the so-called predictor-corrector iteration embedded in the loop for updating the plasma.

\section{Benchmark and example simulations}
\label{sec:example}

In this section, we present comparisons of LWFA related simulations based on QPAD with the PGC and the ones based on full 3D OSIRIS. These benchmark comparisons include the propagation of a Gaussian laser pulse in vacuum and a standard LWFA operating in the nonlinear blowout regime with a matched spot size \cite{Lu2007}.
For these two examples we assume azimuthal symmetry for QPAD and thus only the $m=0$ mode is used. The OSIRIS simulation is full 3D and any physical instabilities or physics that involves higher order modes would be included. We also include a simulation where the LWFA is driven by a Laguerre-Gaussian beam. In the QPAD simulation up to the $m=2$ mode is included. Such a laser driver has been proposed to accelerate positron beams using LWFA \cite{Vieira2014}.

\subsection{Laser pulse propagation in vacuum}

We start by simulating the propagation of a laser pulse with a transverse Gaussian profile (a lowest order Laguerre-Gaussian mode). The laser field evolution has an analytical expression which is the solution to the paraxial Helmholtz equation. The complex amplitude of a certain  $\xi$ slice is given by
\begin{equation}
  a(r,z) = a_0 \frac{w_0}{w(z)}\exp\left(\frac{-r^2}{w(z)^2}\right)
  \exp\left[-\ii\left(\frac{k_0r^2}{2R(z)}-\Psi(z)\right)\right]
\end{equation}
where $z$ (equivalently $s$) is the distance from the focal waist, $w_0$ is the waist radius, $w(z)=w_0\sqrt{1+z^2/z_R^2}$ is the beam radius at $z$, $R(z)=z(1+z_R^2/z^2)$ is the radius of curvature, and $\Psi=\arctan(z/z_R)$ is the Gouy phase. In the benchmark tests, a laser pulse with $k_0w_0=20$ and $a_0=1$ starts from the beam focal waist ($z=0$), and travels a distance of $600k_0^{-1}$, i.e., three Rayleigh lengths, 3 $z_R$. The complex amplitude initially has a longitudinal profile of the form $\cos^2[\pi(\xi-\xi_0)/(2\tau_\text{FWHM})]$ with central position $\xi_0=0$ and  duration $\omega_0\tau_\text{FWHM}=60$. In the OSIRIS simulation, a moving window with spatial resolution of $\Delta_x=\Delta_y=2k_0^{-1}$ and $\Delta_z=0.2k_0^{-1}$ is used. The time step is $\Delta_t=0.125\omega_0^{-1}$. To accurately simulate the group velocity, we used the numerical-dispersion-free field solver \cite{Li2021a} for all the OSIRIS simulations described this paper. In the QPAD simulation, $\Delta_r=1.56k_0^{-1}$ and $\Delta_\xi=0.78k_0^{-1}$. We varied the time steps from $\Delta_s=20\omega_0^{-1}=0.1z_R/c$ to $\Delta_s=200\omega_0^{-1}=z_R/c$.

\begin{figure}[htbp]
\centering
\includegraphics[width=\textwidth]{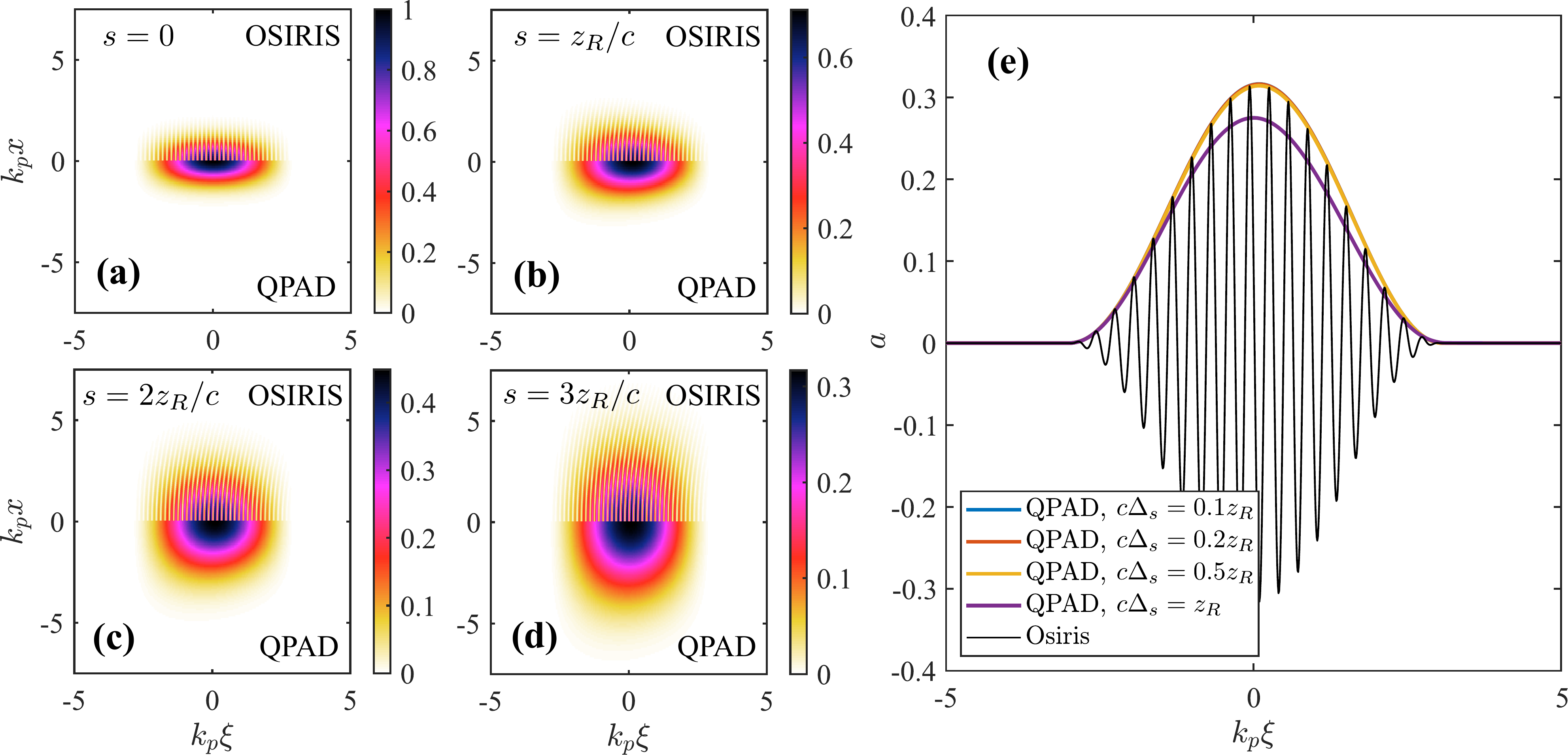}
\caption{Comparison simulations of a freely drifting laser pulse between OSIRIS and QPAD PGC. Snapshots of the laser fields at (a) $s=0$, (b) $s=z_R$, (c) $s=2z_R$ and (d) $s=3z_R$ are presented. The time step of QPAD simulations from (a) to (d) is $\Delta_s=0.2z_R/c$. (e) The impact induced by the time step sizes in QPAD.}
\label{fig:os_benchmark_freestream}
\end{figure}

Fig. \ref{fig:os_benchmark_freestream} summarizes the simulation results. In Fig. \ref{fig:os_benchmark_freestream}(a) to (d), the snapshots of the laser field at four propagation distances ($s$) are presented. The top and bottom halves correspond to the OSIRIS and QPAD simulation results, respectively. Unless otherwise specified, the laser pulse in the simulations here and in the remaining sections moves from the right to left. We carried out a series of simulations to examine the impact of $\Delta_s$. The lineouts at $s=3z_R/c$ for various $\Delta_s$ are shown in Fig. \ref{fig:os_benchmark_freestream}(e). The results for $\Delta_s=0.1,~0.2,~0.5z_R/c$ fully converges to that of OSIRIS while for $\Delta_s=z_R/c$ there is a large deviation. This comparison indicates that the time step needs to properly resolve the evolution of the laser which in vacuum is the Rayleigh length. 

\subsection{LWFA driven by a Gaussian laser pulse}
\label{sec:lwfa}

In this subsection, we provide results of  simulations for a standard LWFA (unloaded) operating in the self guided blowout regime \cite{Lu2007}. In the blowout regime the wake is created by expelling the plasma electrons forward and radially outward where they form a narrow sheath that surrounds a plasma column \cite{Rosenzweig1991,Lu2006}.
In the nonlinear, self-guide LWFA regime a matched spot size
\begin{equation}
\label{eq:match_cond}
  k_p w_0 \simeq 2\sqrt{a_0}
\end{equation}
for the laser is assumed and the laser intensity is relativistic $a_0>1$. Under this condition, the ponderomotive force felt by the plasma background electrons that have been blown out is roughly balanced by the Coulomb force from the ions, and hence the wakefield structure as well as the laser propagation can remain stable without large oscillation in the wake and laser amplitudes. In the simulation, we initialized a laser pulse with $k_pw_0=3.2$, $k_0/k_p=34$ (the Rayleigh length is thus $k_pz_R\simeq 170$), $\omega_p\tau_\text{FWHM}=2.26$ and $a_0=2.5$ in vacuum and it is then focused to the beginning of a plasma density upramp of length $20k_p^{-1}$. We used 8 particles per cell in $r$ and 16 particles distributed in $\phi$. The plasma density was constant at the end of the ramp. In the OSIRIS simulation, the spatial resolution of simulation box is $\Delta_x=\Delta_y=0.1k_p^{-1}$ and $\Delta_z=0.2k_0^{-1}$. The time step is $\Delta_t=0.136\omega_0^{-1}$. In the QPAD simulations $\Delta_r=0.1k_p^{-1}$ and $\Delta_\xi=0.015k_p^{-1}$. Again, only $m=0$ mode is considered due to the assumed azimuthal symmetry.

\begin{figure}[htbp]
\centering
\includegraphics[width=\textwidth]{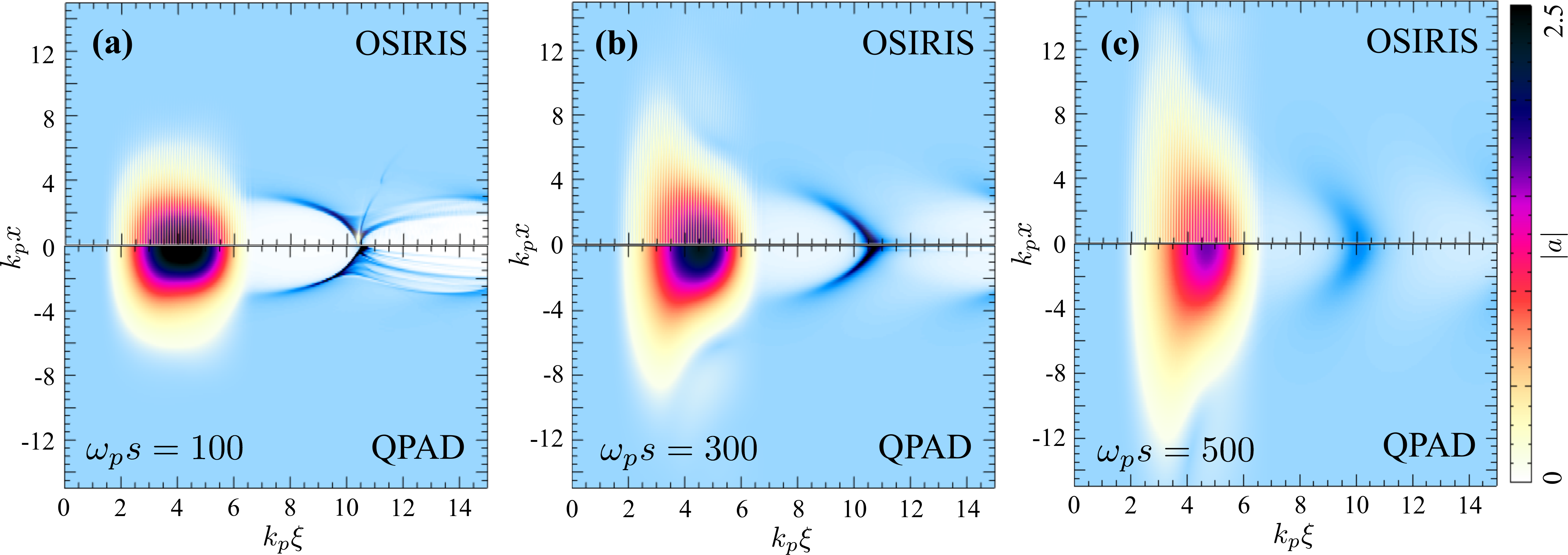}
\caption{Snapshots of the laser fields and background electron density of QPAD and OSIRIS simulations at (a) $s=100\omega_p^{-1}\simeq0.6z_R/c$, (b) $s=300\omega_p^{-1}\simeq1.8z_R/c$, and (c) $s=500\omega_p^{-1}\simeq2.9z_R/c$.}
\label{fig:lwfa1}
\end{figure}

\begin{figure}[htbp]
\centering
\includegraphics[width=\textwidth]{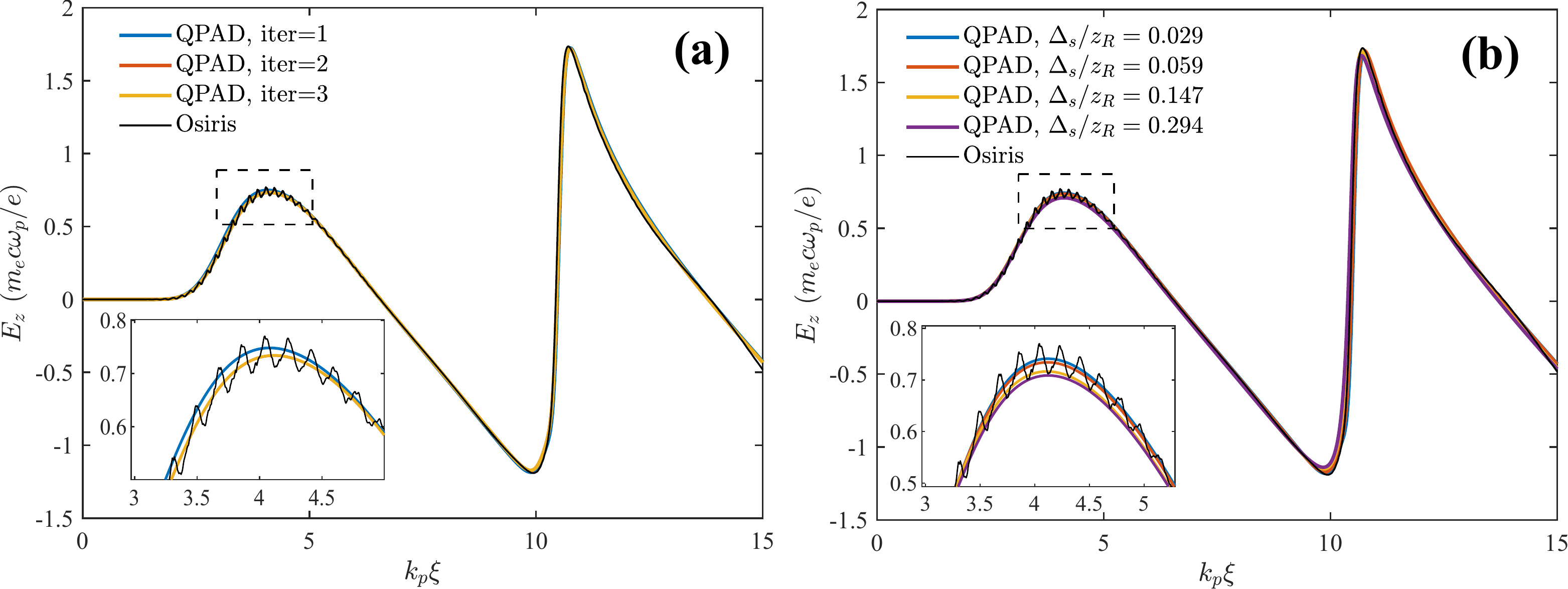}
\caption{Benchmark test results between OSIRIS and QPAD PGC simulations. The impacts induced by (a) the number of iterations for PGC laser solver and (b) the time step sizes in QPAD are also presented.}
\label{fig:lwfa2}
\end{figure}

Fig. \ref{fig:lwfa1}(a) to (c) shows snapshots of the laser field and the background electron density at different propagation distances from both the QPAD and OSIRIS simulations. The time step of the QPAD simulation is $\Delta_s=10\omega_p^{-1}\simeq0.058z_R/c$ and 3 iterations for the PGC laser solver are used. The full set of numerical parameters of QPAD and OSIRIS simulations can be found in Table \ref{tab:config_sim}. It can be clearly seen that the laser diffraction, the erosion of the laser front and the wake shape in both QPAD and OSIRIS agree well with each other.

We compare the impact of the choices for  number of iterations and time step on the wakefield in Fig. \ref{fig:lwfa2} where central lineouts for the axial electric field  $E_z$  are shown. For $\Delta_s=10\omega_p^{-1}\simeq0.058z_R/c$, we can see in Fig. \ref{fig:lwfa2}(a), where $E_z$ at $s=50\omega_p^{-1}$ is shown that the results for using 1, 2 and 3 PGC solver iterations are nearly identical and they also agree well with results from OSIRIS simulations. A closer examination of the $E_z$ field in the dashed box (the inset) indicates that using only one iteration overestimates the $E_z$ amplitude slightly (blue line). In Fig. \ref{fig:lwfa2}(b), results from various time steps but with three iterations for the PGC laser solver are presented.  The results for all time steps, $\Delta_s=0.029,~0.059,~0.147$ and $0.294z_R/c$ are roughly converged and agree well with the OSIRIS result. From the inset in Fig. \ref{fig:lwfa2}(b), it can be seen that the results for the two cases with larger $\Delta_s$ underestimate the wakefield amplitude slightly when compared to the two cases with smaller $\Delta_s$. We also find that increasing the iteration number of the PGC laser solver will not significantly impact the computational performance of simulation. Therefore, in order to guarantee the simulation accuracy, one should avoid using a single iteration for PGC solver and choose a time step that can sufficiently resolve the Rayleigh length.

Particularly noteworthy is the excellent speedup achieved in these QPAD PGC simulations. Both the QPAD and OSIRIS simulations were carried out on the Cori cluster ``Haswell'' nodes at NERSC (Intel Xeon E5-2698 v3 @ 2.30GHz). The numerical configurations and the computation time of both simulations are summarized in Table \ref{tab:config_sim}. Due to the QSA the number of time steps in the QPAD simulation is dramatically decreased compared to the OSIRIS simulation. The azimuthal Fourier decomposition reduces the computational complexity from 3D to 2D, leading to orders of magnitude reduction in the number of cells and particles needed. Therefore, the final core hours consumed by the QPAD simulation is $\sim10^4$ less than that of the OSIRIS simulation. In this comparison we used 16 cores for the QPAD simulation and 2048 for the OSIRIS simulation so in this specific example the wall clock time speedup was $\sim200$. The performance of QPAD-PGC is also evaluated and the detailed scaling test results can be found in the \ref{sec:app_scaling}. For a typical LWFA case, QPAD-PGC can be well scaled to over $10^3$ cores.

\begin{table}[htbp]
\centering
\begin{tabular}{clc}
\hline\hline
\multirow{8}{*}{OSIRIS} & cell sizes $(\Delta_x, \Delta_y, \Delta_z)$ & $0.1~k_p^{-1}$, $0.1~k_p^{-1}$, $5.88\times10^{-3}~k_p^{-1}~(0.2~k_0^{-1})$ \\
                        &  number of cells                            & $300\times300\times2550$ \\
                        &  particles per cell $(N_x, N_y, N_z)$       & 2, 2, 1 \\
                        &  number of particles                        & $9.18\times10^8$ \\
                        &  time step $(\Delta_t)$                     & $0.004~k_p^{-1}~(0.136~k_0^{-1})$ \\
                        &  number of time steps                       & $1.25\times10^4$ \\
                        &  number of processors                       & 2048 \\
                        &  core hours                                 & $\sim 2\times10^4$ hrs. \\ \hline
\multirow{9}{*}{QPAD}   & cell sizes $(\Delta_r, \Delta_\xi)$         & $0.1~k_p^{-1}$, $0.015~k_p^{-1}$ \\
                        &  number of cells                            & $256\times1000$ \\
                        &  particles per cell $(N_r, N_\phi)$         & 8, 16 \\
                        &  number of particles (slice)                & 4096 \\
                        &  number of particles (total)                & $4.096\times10^6$ \\
                        &  time step $(\Delta_t)$                     & $10~k_p^{-1}$ \\
                        &  number of time steps                       & 50 \\
                        &  number of azimuthal mode                   & 1 \\
                        &  number of processors                       & 16  \\
                        &  core hours                                 & 0.8 \\ \hline\hline
\end{tabular}
\caption{\label{tab:config_sim}Configurations and computation time of OSIRIS and QPAD simulations for the LWFA.}
\end{table}

\subsection{LWFA driven by a Laguerre-Gaussian beam}

In this subsection, we demonstrate the capability of the azimuthal mode expansion for modeling physics that requires a finite but small number azimuthal modes via an example where a LWFA is driven by a higher order Laguerre-Gaussian laser pulse. The complex amplitude of the a general Laguerre-Gaussian modes is given by \cite{Allen1992}
\begin{equation}
  a(r,z) = a_0 C_{p,l} \frac{w_0}{w(z)}\left(\frac{r}{w(z)}\right)^{|l|}\exp\left(\frac{-r^2}{w(z)^2}\right) L_p^{|l|}\left(\frac{2r^2}{w^2(z)}\right)
  \exp\left[-\ii\left(\frac{k_0r^2}{2R(z)}+l\phi-\Psi(z)\right)\right]
\end{equation}
where $p$ and $l$ are the radial and azimuthal indices, $L_p^{|l|}$ is the generalized Laguerre polynomial, $C_{p,l}$ is the normalized amplitude of the mode, and the definitions for $w(z)$ and $R(z)$ are identical to those of the fundamental Gaussian mode. 
The Gouy phase shift of a Laguerre-Gaussian beam is exaggerated by the factor $2p+|l|$, i.e., $\Psi(z)=(2p+|l|+1)\arctan(z/z_R)$. For the modes $p=0$ and $|l|>0$, the laser intensity vanishes at $r=0$, and increases and then decreases radially, presenting a donut-like intensity distribution. The associated ponderomotive force will push the background electrons inside the donut (the region where the $\PP{|a|^2}{r}$ is positive) inward, which  then forms a high-density electron column on the axis. This particular field structure can provide a focusing force for positive charged particles and hence has been considered for positron beam acceleration \cite{Vieira2014}.

\begin{figure}[htbp]
  \centering
  \includegraphics[width=\textwidth]{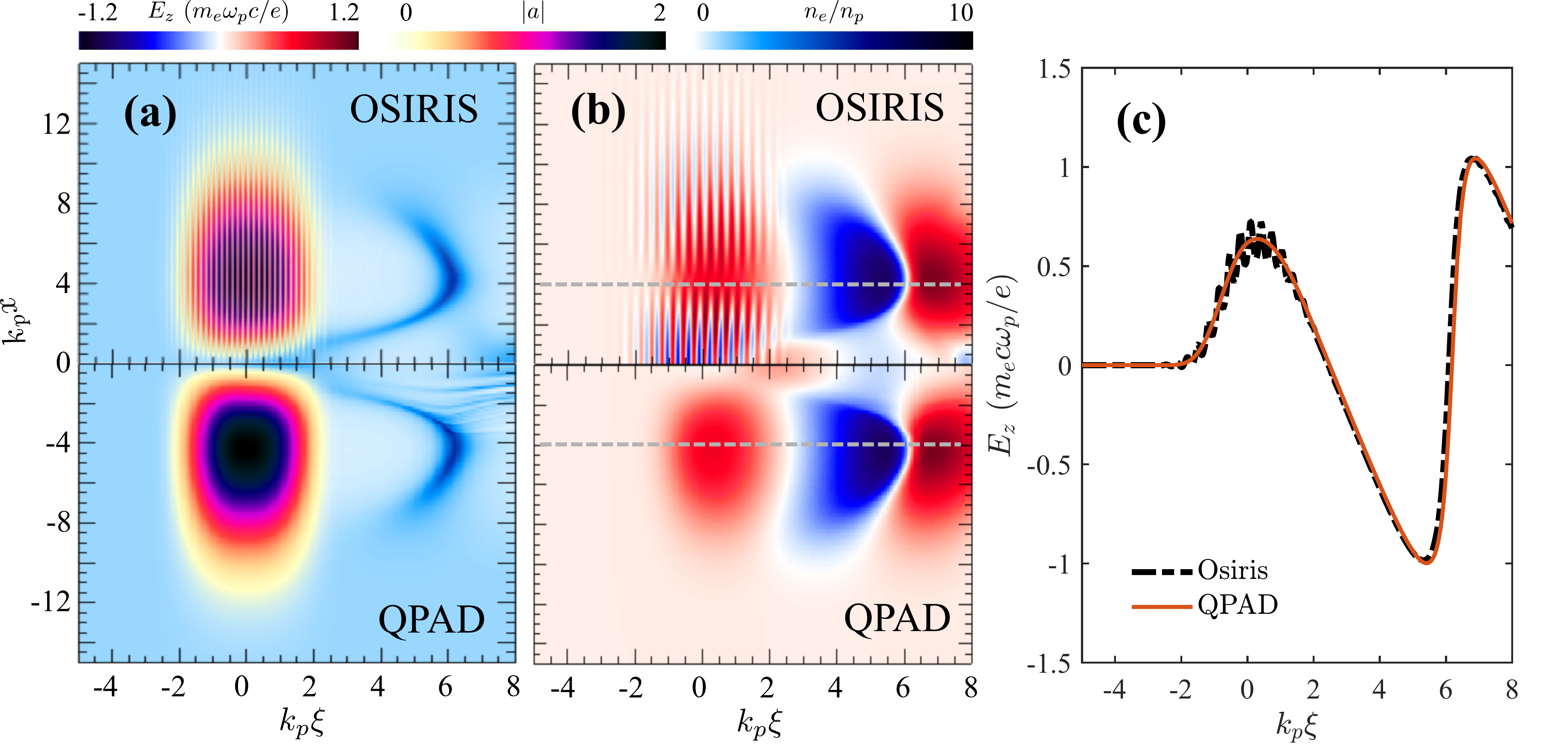}
  \caption{QPAD-PGC simulation of a LWFA driven by a Laguerre-Gaussian laser pulse with $l=1$ and $p=0$. Snapshots of (a) the laser fields and plasma electron density, (b) axial field $E_z$ and (c) lineouts of $E_z$ at $k_pr=4$ are taken at $\omega_ps=40$.}
  \label{fig:oam}
\end{figure}

Fig. \ref{fig:oam} shows the plasma wake excited by a laser pulse with $(p,l)=(0,1)$, $a_0=2.1$, $k_pw_0=6$, $\omega_0/\omega_p=20$ and $\omega_p\tau_\text{FWHM}=2.3$. The simulation box has a dimension of $25.6k_p^{-1}\times 13k_p^{-1}$ in the radial and axial directions and the spatial resolution is $\Delta_r=0.1k_p^{-1}$ and $\Delta_\xi=0.04 k_p^{-1}$. The time step is $\Delta_s=5\omega_p^{-1}$. To well resolve the asymmetry due to the $e^{-\ii l\phi}$ term, the azimuthal mode expansion is truncated at $m=2$. Figs. \ref{fig:oam}(a) to (c) present the background electron density, axial field $E_z$ at $\omega_ps=40$ and the lineouts of $E_z$ at $k_pr=4$, respectively. In Fig. \ref{fig:oam}(a) and (b), the top and bottom subfigures correspond to the results from 3D OSIRIS and QPAD-PGC simulations, respectively. As expected, the laser with a donut-shaped intensity profile simultaneously repels the ambient electrons inward and outward, forming an electron-rarefied toroidal wake behind. From the Fig. \ref{fig:oam}(c), it is clear that the lineout of $E_z$ at $k_pr=4$ [denoted by the dashed lines in Fig. \ref{fig:oam}(b)] from the QPAD simulation is in excellent agreement with that from a full 3D OSIRIS simulation.

\section{Conclusion}

In summary, we have described a new quasi-static based particle-in-cell algorithm  that utilizes the ponderomotive guiding center approach together with an azimuthal mode expansion.
The proposed algorithm is not restricted by the regular CFL condition for the time step as is the case for a standard fully electromagnetic PIC code. The algorithm can also be asynchronously parallelized (pipelined). The algorithm has been implemented into the quasi-static PIC code QPAD. In QPAD, the laser envelope equation, the Maxwell's equations and all the physical quantities are decomposed into azimuthal harmonics, i.e., Fourier modes.
With this hybrid approach where a PIC description is used in $r-z$ and a gridless description is used in $\phi$, QPAD with PGC model can more efficiently simulate short pulse laser-plasma interactions than full 3D PIC, quasi-3D explicit PIC, or full 3D quasi-static codes. The implementation of the plasma particle pusher and the plasma susceptibility deposition are also described. This new algorithm was benchmarked and compared against the results from the full 3D PIC code OSIRIS for a few sample cases. Excellent agreement was achieved for the simulations of a laser propagating freely in vacuum and a standard LWFA problem. Enormous reduction of in core-hours ($\sim$4 orders of magnitude) compared to a full 3D PIC code were obtained while maintaining high fidelity. We also simulated a LWFA driven by a Laguerre-Gaussian laser pulse as an example to show the capability of multi-mode simulations. With the PGC algorithm, QPAD can also very effectively model PWFA and LWFA with trailing particle beams and photon acceleration. Future directions for algorithm enhancement will include mesh refinement and improving the envelope solver to handle pump depletion distances.  

\section*{Acknowledgments}

This work was supported in parts by the US Department of Energy through SciDAC FNAL subcontract 644405 and contract number DE-SC0010064 and US National Science Foundation grant numbers 2108970. The work of W.A. was supported by the National Natural Science Foundation of China (NSFC) grant number 12075030. Simulations were carried out on the Cori Cluster of the National Energy Research Scientific Computing Center (NERSC).

\begin{appendix}

\section{Derivation of iterative convergence conditions for a uniform $\chi^n$}
\label{sec:app_iter_conv_cond}

Given a simulation time step $s=n\Delta_s$, the quantities $a_{j-1}^{n+\hf}$, $a_{j-2}^{n+\hf}$  and $a_j^{n-\hf}$ in Eq. (\ref{eq:fd_iter}) can be viewed as constants since here we only explore how the solution converges as increasing $l$. Subtracting Eq. (\ref{eq:fd_iter}) for $l$ and $l-1$ gives,
\begin{equation}
  \left(\ii k_0 - \frac{1}{4}\Delta_s\nabla_\perp^2 + \frac{3}{2\Delta_\xi}\right)r^l
  = \frac{1}{4}\Delta_s\chi^n r^{l-1}
\end{equation}
where  $r^l\equiv a_j^{n+\hf,l}-a_j^{n+\hf,l-1}$ and $\chi^n$ is assumed to have no spatial dependence. We define $\ii[\mathbf{k}_\perp]$ as the counterpart to $\nabla_\perp$ in the $\mathbf{k}$-space such that in the Cartesian coordinates, $[\mathbf{k}_\perp]$ is given by
\begin{equation}
  [\mathbf{k}_\perp] = \frac{\sin(k_x\Delta_x/2)}{\Delta_x/2} \mathbf{e}_x + \frac{\sin(k_y\Delta_y/2)}{\Delta_y/2} \mathbf{e}_y,
\end{equation}
where $\Delta_x,~\Delta_y$ are the cell sizes of the mesh-grid in the $x$-$y$ plane.
Using the ansatz $r^l = \tilde{r}^l e^{\ii \mathbf{k}_\perp\cdot\mathbf{x}_\perp}$, the ratio of the residuals of two adjacent iterations is given by
\begin{equation}
  \left|\frac{\tilde{r}^l}{\tilde{r}^{l-1}}\right|^2 = \frac{\frac{1}{16}\Delta_s^2|\chi^n|^2}{\left(\frac{1}{4}\Delta_s[k_\perp]^2 + \frac{3}{2\Delta_\xi}\right)^2 + k_0^2} \le 
  \frac{\frac{1}{16}\Delta_s^2|\chi^n|^2}{\frac{9}{4\Delta_\xi^2} + k_0^2},
\end{equation}
A necessary condition for convergence is $|\tilde{r}^l/\tilde{r}^{l-1}|<1$ which leads to
\begin{equation}
  k_0^2\Delta_\xi^2 + \frac{9}{4} > \frac{1}{16}|\chi^n|^2\Delta_s^2\Delta_\xi^2.
\end{equation}

\section{Proof of numerical stability for a uniform $\chi^n$ and $l\rightarrow\infty$}
\label{sec:app_numerical_stability}

We carry out a von Neumann analysis to show that the proposed laser solver is unconditionally stable when $\chi$ has no spatial dependence for each time step and the number of iterations approaches infinity. In this situation, the $a^{n+\hf,l}$ and $a^{n+\hf,l-1}$ converge to $a^{n+\hf}$ and Eq. (\ref{eq:fd_iter}) actually becomes Eq. (\ref{eq:fd_direct}). The round-off error $\epsilon^n={a}_N^n-a^n$ ($a_N^n$ is the numerical solution obtained in finite precision arithmetic and $a^n$ is precise solution of the difference equation) still satisfies Eq. (\ref{eq:fd_direct}) in the sense of machine precision. Using the ansatz $\epsilon^n=g^n e^{\ii k_\xi\xi} e^{\ii\mathbf{k}_\perp\cdot\mathbf{x}_\perp}$ where $g$ is the error growth factor, we can obtain
\begin{equation}
  \left(\ii k_0 + \frac{1}{4}\Delta_s ([k_\perp]^2-\chi^n) \right)\epsilon^{n+\hf} + \ii[k_\xi]\epsilon^{n+\hf} =
  \left(\ii k_0 - \frac{1}{4}\Delta_s ([k_\perp]^2-\chi^n) \right)\epsilon^{n-\hf} + \ii[k_\xi]\epsilon^{n-\hf}
\end{equation}
where  $\ii[k_\xi]$ is the counterpart of $\PP{}{\xi}$ in the $\mathbf{k}$-space. For the backward difference operator in $\xi$ used in this paper, $[k_\xi]$ is given by
\begin{equation}
\label{eq:k_xi_brac}
  [k_\xi] = \left( \frac{3}{2}e^{-\ii\hf k_\xi\Delta_\xi} - \frac{1}{2}e^{-\ii\frac{3}{2} k_\xi\Delta_\xi} \right)
  \frac{\sin(k_\xi\Delta_\xi/2)}{\Delta_\xi/2}.
\end{equation}
The growth rate of the numerical error is
\begin{equation}
  g = \frac{\epsilon^{n+\hf}}{\epsilon^{n-\hf}} = 
  \frac{\ii(\mathfrak{Re}\{[k_\xi]\}+k_0) - \mathfrak{Im}\{[k_\xi]\} - \frac{1}{4}\Delta_s ([k_\perp]^2-\chi^n)}
  {\ii(\mathfrak{Re}\{[k_\xi]\}+k_0) - \mathfrak{Im}\{[k_\xi]\} + \frac{1}{4}\Delta_s ([k_\perp]^2-\chi^n)}.
\end{equation}
According to Eq. ({\ref{eq:k_xi_brac}}) the imaginary part of $[k_\xi]$
\begin{equation}
  \mathfrak{Im}\{[k_\xi]\} = \left[\frac{1}{2}\sin\left(\frac{3}{2}k_\xi\Delta_\xi\right)
  -\frac{3}{2}\sin\left(\frac{1}{2}k_\xi\Delta_\xi\right)\right]\frac{\sin(k_\xi\Delta_\xi/2)}{\Delta_\xi/2} < 0
\end{equation}
for any real $k_\xi$. Since $\chi^n < 0$ according to Eq. ({\ref{eq:chi_def}}), $[k_\perp]^2-\chi^n > 0$. Therefore,
\begin{equation}
  |g|^2 = 
  \frac{\left(\mathfrak{Re}\{[k_\xi]\}+k_0\right)^2 + \left(\frac{1}{4}\Delta_s ([k_\perp]^2-\chi^n) + \mathfrak{Im}\{[k_\xi]\}\right)^2}
  {\left(\mathfrak{Re}\{[k_\xi]\}+k_0\right)^2 + \left(\frac{1}{4}\Delta_s ([k_\perp]^2-\chi^n) - \mathfrak{Im}\{[k_\xi]\}\right)^2} \leq 1
\end{equation}
and the equality holds only when $k_\xi=0$. $|g|\leq1$ means the round-off error will not grow as $n$ increases and the algorithm is thus numerically stable. This conclusion does not depend on the specific forms of $[\mathbf{k}_\perp]$ which implies that the numerical stability is independent of the coordinate system and specific form of the difference operator in the transverse plane.

\section{Parallel scaling}
\label{sec:app_scaling}

Like the original QPAD, QPAD-PGC is also parallelized using MPI and runs on distributed memory systems. In this appendix, we present the scalings of QPAD-PGC on the Cori cluster at NERSC. For a QSA code, the longitudinal and transverse parallel scalability differ from each other due their use of fundamentally different parallelization schemes. In the longitudinal direction the code is parallelized using a pipelining algorithm which has shown excellent scalability in previous work \cite{Feng2009}. In the transverse direction the code is parallelized using domain decomposition and the parallelization does not scale linearly since the field and laser solvers use the cyclic reduction method. To quantify the parallel scalability in the different directions, we carried out  transverse, longitudinal and bi-directional scaling tests. In the transverse and longitudinal tests, the simulation window is partitioned only in the transverse and longitudinal directions respectively, while in the bi-directional test the window is partitioned in both directions and the number of cores in both directions are set to be equal.

\begin{figure}[htbp]
  \centering
  \includegraphics[width=\textwidth]{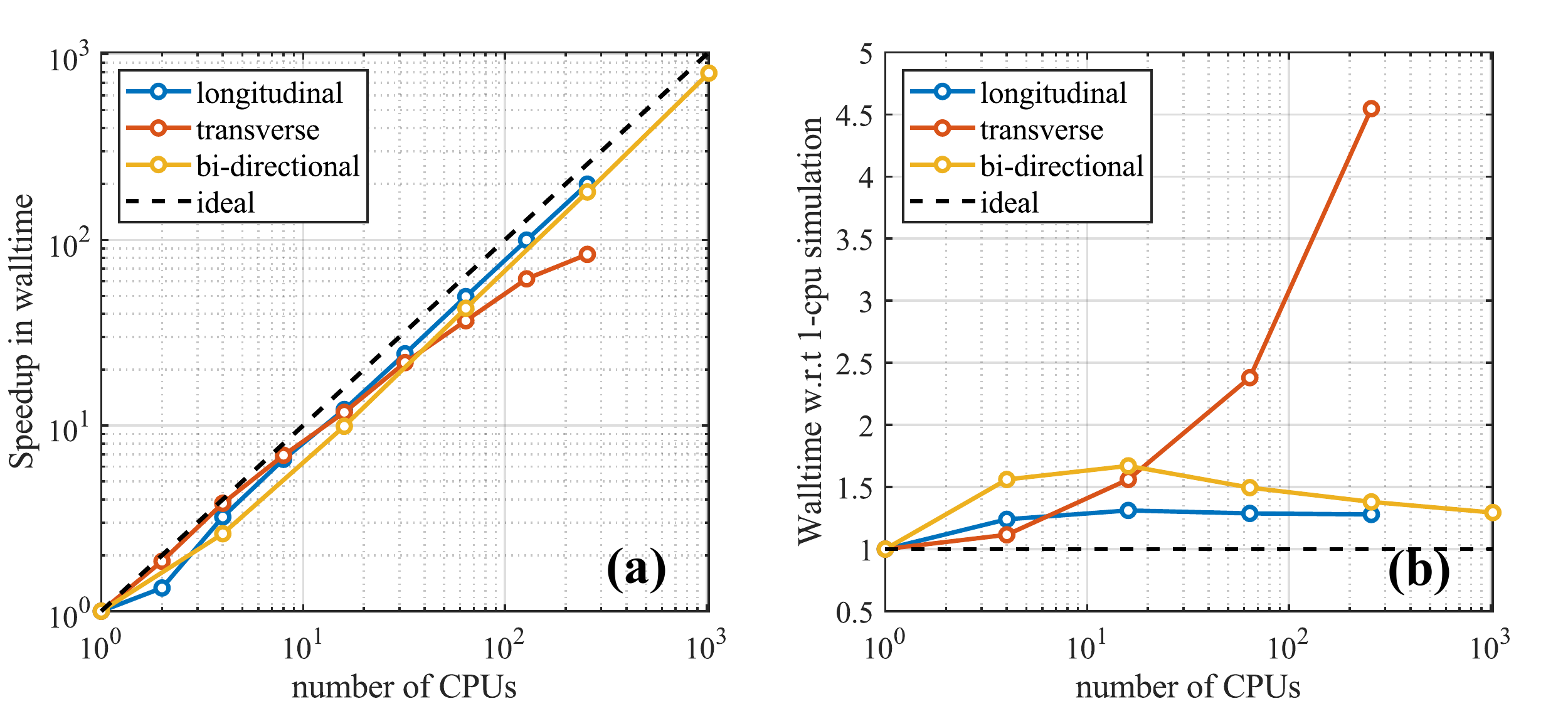}
  \caption{(a) Strong scaling and (b) weak scaling of QPAD-PGC.}
  \label{fig:scaling}
\end{figure}

An LWFA simulation with the same physical parameters as Section \ref{sec:lwfa} is used for the scaling tests. In the strong scaling test the total problem size is fixed and the relation between the runtime and the number of cores is examined. The simulation window has $2^{14}\times2^{13}$ cells in the transverse and longitudinal directions, and there are 8 macro-particles within a radial cell size $\Delta_r$ and 8 macro-particles distributed in $2\pi$ angle. In the weak scaling test, the same physical problem is simulated but the resolution is varied from $2^9\times2^8$ cells to $2^{14}\times2^{13}$ cells where the numbers of MPI partitions are varied from 1 to 1,024 respectively. Thus, each MPI partition is fixed with $2^{17}$  cells (the partition shape may vary for different total number of cores) in each case and the same number of particles per cell as for the strong scaling test is used. The results of the scaling benchmarks are presented in Fig. \ref{fig:scaling}. Due to the excellent scalability of pipelining algorithm, the pure longitudinal partitioning (blue lines) in both strong and weak scaling tests are very close to the ideal performance (dashed line). For the strong scaling the pure transverse partitioning (red lines) start to deviate significantly from the ideal performance when the number of cores is roughly larger than 100 (or the number of cells per core becomes less than $\sim10^6$). The performance of the bi-directional partitioning (yellow lines) lies between the pure transverse and longitudinal partitioning when the number of cores $\gtrsim20$. The one-step single-core run of the strong scaling test ($2^{14}\times2^{13}$ cells) costs 3300 core seconds. All the above tests excluded the file I/O.

\end{appendix}

\bibliographystyle{elsarticle-num}
\bibliography{refs.bib}

\end{document}